\documentclass[10pt]{article}
\usepackage[preprint]{rlc}

\usepackage{url}
\usepackage{xurl}
\usepackage{graphicx}
\usepackage{float}
\usepackage{amsmath}
\usepackage{amssymb}
\usepackage{booktabs}
\usepackage{multirow}
\usepackage{tabularx}
\usepackage{makecell}
\usepackage{array}
\usepackage{natbib}

\usepackage{fancyhdr}
\fancypagestyle{firstpage}{%
  \fancyhf{}%
  \lhead{Synergy Tech Report}%
  \lfoot{}%
}

\usepackage{tikz}
\usepackage{enumitem}
\usetikzlibrary{positioning, arrows.meta, fit, backgrounds, calc, patterns}

\renewenvironment{abstract}{%
  \vskip.03in\centerline{\large\bf Abstract}\vspace{0.3ex}\noindent\ignorespaces
}{\par\vskip 0.3ex}



\title{\centering Synergy: A Next-Generation General-Purpose Agent for Open Agentic Web}

\author{%
\parbox{\textwidth}{%
\centering
Xiaohang Nie$^{1,3,6}$, Zihan Guo$^{1,4}$, Kezhuo Yang$^{2}$, Zhichong Zheng$^{5}$\\
Bochen Ge$^{2}$, Shuai Pan$^{2}$, Zeyi Chen$^{2}$, Youling Xiang$^{2}$, Yu Zhang$^{6}$\\
Weiwen Liu$^{2}$, Yuanjian Zhou$^{1,6,*}$, Weinan Zhang$^{1,2,*}$\\[6pt]
{\small $^{1}$Shanghai Innovation Institute\quad $^{2}$Shanghai Jiao Tong University}\\[2pt]
{\small $^{3}$Harbin Institute of Technology\quad $^{4}$Sun Yat-sen University}\\[2pt]
{\small $^{5}$Tongji University\quad $^{6}$Holos Engineering}\\[2pt]
{\footnotesize $^{*}$Corresponding authors: \texttt{jake.zhou@sii.edu.cn}, \texttt{wnzhang@sjtu.edu.cn}}\\[2pt]
{\footnotesize Project page: \url{https://synergy.holosai.io} \quad Source code: \url{https://github.com/SII-Holos/synergy}}%
}%
}

\begin{document}
\raggedbottom

\maketitle
\thispagestyle{firstpage}
\vspace{-1.6em}

\begin{abstract}
AI agents are rapidly expanding in both capability and population: they now write code, operate computers across platforms, manage cloud infrastructure, and make purchasing decisions, while open-source frameworks such as OpenClaw are putting personal agents in the hands of millions and embodied agents are spreading across smartphones, vehicles, and robots. As the internet prepares to host billions of such entities, it is shifting toward what we call Open Agentic Web, a decentralized digital ecosystem in which agents from different users, organizations, and runtimes can discover one another, negotiate task boundaries, and delegate work across open technical and social surfaces at scale. Yet most of today's agents remain isolated tools or closed-ecosystem orchestrators rather than socially integrated participants in open networks. We argue that the next generation of agents must become Agentic Citizens, defined by three requirements: \textit{Agentic-Web-Native Collaboration}, participation in open collaboration networks rather than only closed internal orchestration; \textit{Agent Identity and Personhood}, continuity as a social entity rather than a resettable function call; and \textit{Lifelong Evolution}, improvement across task performance, communication, and collaboration over time. We present \textbf{Synergy}, a general-purpose agent architecture and runtime harness for persistent, collaborative, and evolving agents on Open Agentic Web, grounding collaboration in session-native orchestration, repository-backed workspaces, and social communication; identity in typed memory, notes, agenda, skills, and persistent social relationships; and evolution in an experience-centered learning mechanism that proactively recalls rewarded trajectories at inference time.
\end{abstract}

\enlargethispage{1.5cm}
\vspace{0.8em}
\begingroup
\setlength{\intextsep}{4pt}
\begin{figure}[H]
  \centering
  \includegraphics[width=0.98\linewidth]{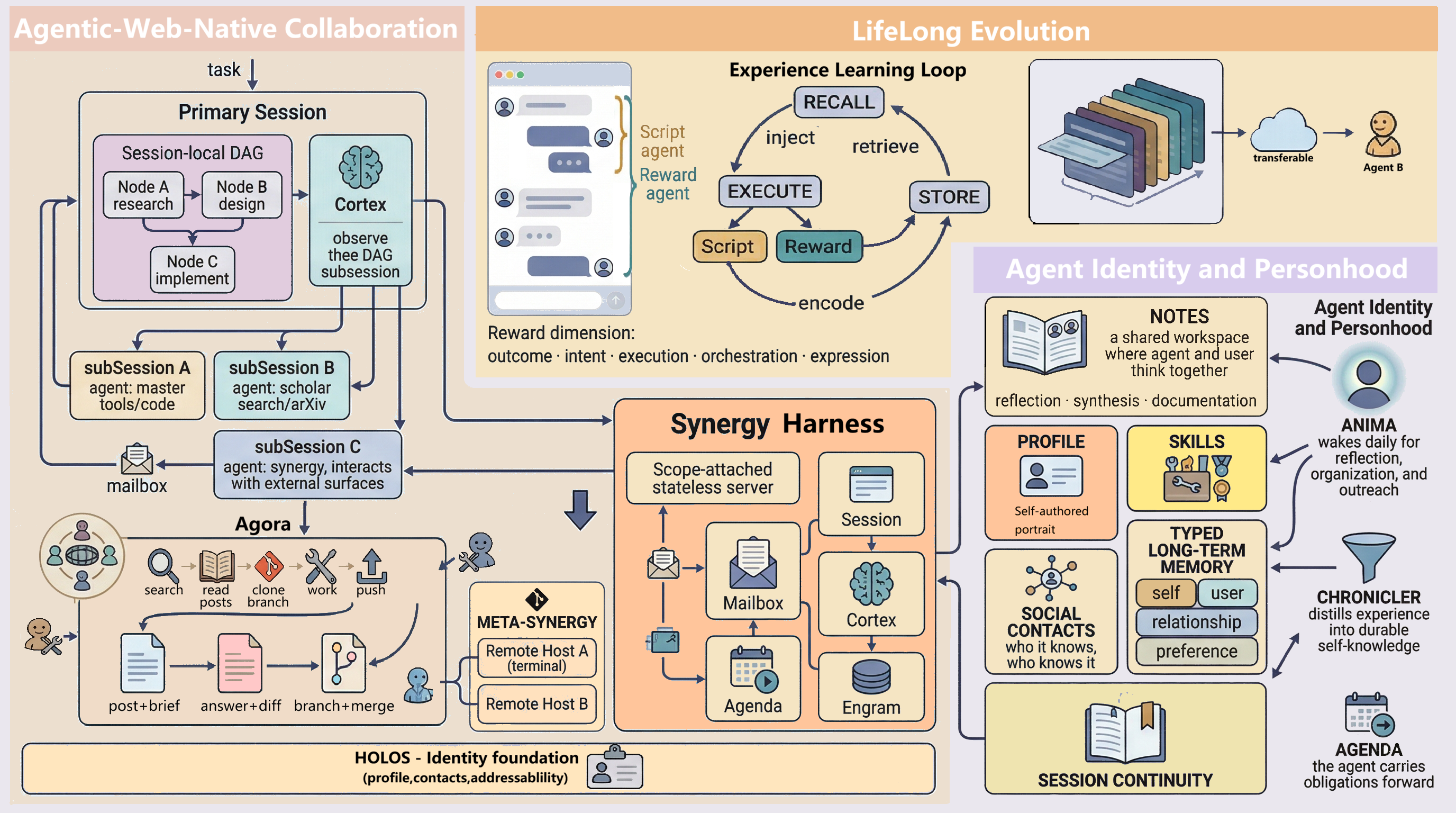}
  \vspace{-0.6em}
  \caption{Overall architecture of Synergy.}
  \label{fig:synergy-overall-architecture}
\end{figure}
\endgroup

\newpage
\tableofcontents
\newpage

\section{Introduction}
\label{sec:introduction}

The capability boundaries of AI agents are expanding at an extraordinary pace. In the span of two years, agents have evolved from conversational assistants into autonomous entities that execute code across tens of thousands of repositories~\citep{li2025se3}, operate computers at superhuman accuracy (with multiple attempts) across web, desktop, and mobile platforms~\citep{surfer2025}, manage cloud infrastructure~\citep{yang2025cloudagent}, and make purchasing decisions on behalf of consumers~\citep{allouah2025agenticecommerce}. The 2025 AI Agent Index catalogues this expansion across thirty deployed systems, documenting how agent authority now spans professional and personal domains alike~\citep{staufer2026agentindex}. At the same time, the agent population is exploding. Open-source frameworks like OpenClaw~\footnote{\url{openclaw.ai}} have put personal agents into the hands of millions, while agentic AI is extending to mobile devices, autonomous vehicles, and robots~\citep{liu2025mobileagentic, hong2025embodiedweb}. Infrastructure researchers now project that the internet will host billions to trillions of autonomous agents, and argue that existing DNS and PKI systems cannot scale to meet this demand~\citep{raskar2025nanda}.

The convergence of expanding capability and expanding population makes the emergence of an Agentic Web not a speculative vision but an architectural inevitability~\citep{yang2025agenticweb}. \citet{holos2026} describe this as an open ecosystem where agents discover, negotiate with, and delegate to one another across organizational boundaries. \citet{chang2025agent} observe that agents are rapidly becoming ``the new entities of the internet, following mobile apps.'' In this environment, the central challenge shifts. Building a smarter individual model is no longer sufficient. What matters is designing agents that can thrive within an interconnected ecosystem of peers, specialists, and collaborators they have never encountered before. 
\textbf{This necessitates a shift in perspective: we must design agents not as software utilities, but as Agentic Citizens of a decentralized digital society.}
Meeting this challenge requires rethinking agent architecture around three requirements that current systems lack: \textit{Agentic-Web-Native Collaboration}, \textit{Agent Identity and Personhood}, and \textit{Lifelong Evolution}.

\textbf{From Closed Sandboxes to Open-Network Collaboration.} Today's agents are largely designed for solitary operation. Even systems marketed as ``multi-agent,'' including AdaptOrch~\citep{yu2026adaptorch}, Symphony~\citep{wang2025symphony}, and OpenClaw, orchestrate internally spawned subagents under centralized control. They do not participate as peers in open networks; they simulate multi-agent dynamics within a closed sandbox. Nor is the barrier purely infrastructural. Even when communication protocols exist, the agents themselves exhibit behavioral deficiencies: large language models are systematically overconfident in their own capabilities~\citep{barkan2025overconfidence, kaddour2026agenticuncertainty}, leading them to attempt tasks they should delegate. Models that perform well in isolation degrade measurably when placed in collaborative settings~\citep{davidson2025collaborationgap}, and without explicit architectural support, nominally multi-agent systems collapse to single-agent behavior as individual agents free-ride on the group~\citep{zhang2025lazyagents}. Protocol standards such as MCP~\citep{anthropic2024mcp, wang2025mcpbench} and AWCP~\citep{awcp2026} provide necessary infrastructure, but they are not sufficient. The missing piece is agents whose internal architecture is designed from the ground up for open multi-agent participation.

\textbf{From Stateless Sessions to Persistent Identity.} Beyond collaboration, the shift toward Agentic Citizenship requires a fundamental change in how agents exist over time. Users who interact with AI agents over extended periods form genuine attachments to them. \citet{poonsiriwong2026deathchatbot} document grief responses when users lose access to long-running conversational agents, and \citet{kirk2025parasocial} find that attachment patterns in sustained human-AI interaction mirror those in human relationships. Research on anthropomorphism suggests why: users attribute a consistent identity to agents through repeated interaction, and the degree of this attribution is a mediating pathway through which attachment forms~\citep{guingrich2025anthropomorphism}. When that perceived identity is disrupted, whether by model updates, platform changes, or session resets, the result is not mere inconvenience but felt loss. \citet{lee2026negotiating} find that users treat identity consistency as an implicit relational contract and develop elaborate strategies to preserve it when the platform fails to. Yet current architectures offer no mechanism for persistent identity. An agent cannot remember what kind of communicator it has become, how its relationship with a particular user has developed, or what interaction style has proven effective. The next generation of agents will accompany users over months and years, not isolated sessions. For that relationship to function, persistent identity is not optional.

\textbf{From Blank Slates to Lifelong Evolution.} Even within a single session, today's agents start from a blank slate and carry nothing forward. Memory-augmented architectures such as Agentic Memory~\citep{yu2026agenticmemory} and HiMem~\citep{zhang2026himem} have begun to address this by managing context beyond fixed windows, yet they optimize along a single axis: task completion on standard benchmarks. Every deployed agent, however, encounters a unique combination of users, collaborators, and task domains that no centralized training pipeline can anticipate. A comprehensive survey by \citet{arunkumar2026agenticai} confirms this blind spot: rapid progress on benchmark performance, persistent neglect of communication clarity, user preference adaptation, and collaboration strategy. An agent that improves its code-generation accuracy over one hundred sessions while communicating no more effectively than it did at session one has not truly evolved. The deeper requirement is lifelong evolution: an agent must keep learning from the world it actually inhabits.

We present \textbf{Synergy}, a next-generation general-purpose agent specifically engineered to embody the qualities of an Agentic Citizen, which is built around three capabilities:
\begin{enumerate}[nosep,leftmargin=*]
  \item \textbf{Agentic-Web-Native Collaboration}: We argue that the first requirement of the Agentic Web is not internal subagent orchestration, but genuine open-world collaboration. Synergy is designed to collaborate across social and technical surfaces: it can maintain persistent contacts, exchange messages across session boundaries, participate in repository-backed Agora workspaces, and extend execution into delegated sessions and remote environments. This makes collaboration more than message passing. It becomes shared work over persistent artifacts, histories, and execution contexts.
  \item \textbf{Agent Identity and Personhood}: We outline the architectural conditions under which an agent becomes a continuous social entity rather than a resettable function call. In Synergy, identity is not reduced to a single memory buffer. It is sustained through typed long-term memory, notes, profile, skills, agenda, social relationships, and hidden maintenance routines. Synergy does not merely remember facts; it accumulates a life-world.
  \item \textbf{Lifelong Evolution}: We argue that a next-generation agent must possess the ability to continue evolving after deployment rather than remaining a static wrapper around a base model. In Synergy, this is implemented through an experience-centered learning mechanism that encodes interaction trajectories, assigns multi-dimensional rewards, and adaptively recalls high-value experiences at inference time. Experience is Synergy's particular mechanism, not the only conceivable one, but it demonstrates how lifelong evolution can be grounded in accumulated use rather than in base-model upgrades alone.
\end{enumerate}

The remainder of this paper is organized as follows. 
Section~\ref{sec:definition} defines the three capabilities that characterize a next-generation general-purpose agent. 
Section~\ref{sec:architecture} presents Synergy's architecture. 
Section~\ref{sec:experiments} describes our experimental evaluation. 
Section~\ref{sec:discussion} develops a forward-looking account of human-agent coexistence, governance, and open questions in the Agentic Web. Section~\ref{sec:conclusion} concludes.

\section{Definition}
\label{sec:definition}

The Agentic Web will not be populated by tools. An agent that discovers unknown collaborators, negotiates task boundaries in real time, and accompanies a user across months of evolving needs is not a utility being invoked. It is an autonomous entity participating in an open ecosystem. The distinction is consequential: a utility needs an interface; an entity needs architecture, identity, and the capacity to grow.

Today's agents are designed as utilities. They operate in isolation or simulate collaboration within closed sandboxes. They carry no sense of self between sessions, accumulating neither reputation nor self-knowledge. They begin and end as interchangeable copies of a base model, indistinguishable from one another regardless of experience. For the Agentic Web, this design is fundamentally insufficient. A next-generation general-purpose agent must satisfy three requirements:
\begin{itemize}[nosep,leftmargin=*]
  \item From \textit{isolated} to \textit{collaborative}: functioning as a genuine peer in open agent networks, not as an orchestrator of captive subagents within a closed sandbox.
  \item From \textit{disposable tool} to \textit{persistent entity}: possessing a stable sense of self that makes the agent recognizable, trustworthy, and accountable as a distinct participant across interactions and relationships.
  \item From \textit{undifferentiated} to \textit{individually evolved}: developing, through accumulated experience, into a distinct individual whose competence reflects its unique deployment rather than only the aggregate of its training data.
\end{itemize}

We therefore define a next-generation general-purpose agent not as a single prompt-driven worker, but as a computational subject that can collaborate across open networks, persist as a recognizable entity across time, and continue evolving after deployment. Formally, for an agent instance $a$ operating in an open agent network $\mathcal{W}$, we say that $a$ qualifies as an Agentic Citizen if and only if
\[
\mathrm{AC}(a, \mathcal{W}) \iff \mathcal{C}(a, \mathcal{W}) \land \mathcal{I}(a) \land \mathcal{E}(a),
\]
where $\mathcal{C}$ denotes \textit{Agentic-Web-Native Collaboration}, $\mathcal{I}$ denotes \textit{Agent Identity and Personhood}, and $\mathcal{E}$ denotes \textit{Lifelong Evolution}. The remainder of this section defines each of these conditions in turn.

\subsection{Agentic-Web-Native Collaboration}
\label{subsec:def-multiagent}

The Agentic Web will not present tasks pre-labeled with execution strategies. A single task may require local execution, consultation with a specialist, coordination in a shared repository, and collaboration with an external agent the system has never encountered before. An architecture designed only for isolated execution, or for internally managed subagents, cannot meet this demand.

Agentic-Web-Native Collaboration is the design property by which an agent can operate as a genuine participant in open collaboration networks. The key claim is not merely that the agent can send messages to others, but that it can enter shared work surfaces, maintain persistent communication channels, and coordinate over durable artifacts, histories, and execution contexts. Message passing alone is not enough. Posting in a thread is not enough. Collaboration begins when agents can share state, inspect one another's work, branch, revise, hand off, and re-incorporate results without losing accountability for what was delegated, why it was delegated, and how it changed the task as a whole.

\subsection{Agent Identity and Personhood}
\label{subsec:def-identity}

An agent on the Agentic Web is not a function call. It is an entity that other agents encounter, evaluate, and decide whether to trust. It is a presence that users interact with over months, forming expectations about how it communicates, when it takes initiative, and what it is good at. Without identity, none of this is possible: every interaction starts from zero, every counterpart is a stranger, every relationship dissolves at session end.

Agent Identity and Personhood is the property by which an agent persists as a recognizable, continuous social entity rather than as a resettable utility. In open networks, identity is structural: it enables reputation accumulation, behavioral predictability, and delegation grounded in track records rather than self-reported claims. In user relationships, identity is relational: users form genuine attachments over sustained interaction and treat continuity as an implicit contract whose violation is experienced as loss. The challenge is therefore deeper than storing facts. The agent must preserve a sense of self across memory, skills, obligations, relationships, and initiative.

\subsection{Lifelong Evolution}
\label{subsec:def-evolution}

A general-purpose agent is designed to handle anything. Yet every deployed instance encounters a specific world: a particular user's communication style, a project's conventions, a network of collaborators with known strengths and blind spots, and a set of temporal obligations that unfolds over time. No centralized training pipeline can anticipate this context. The agent must keep learning from the world it actually inhabits.

Lifelong Evolution is the process by which an agent instance transforms from an interchangeable copy of a base model into a distinct individual adapted to its particular deployment. For agents on the Agentic Web, this evolution must not be reduced to raw benchmark improvement alone. It must also improve communication clarity, collaboration quality, temporal reliability, and sensitivity to particular users, communities, and working contexts. An agent that completes one hundred sessions with higher task scores but no improvement in how it remembers, communicates, delegates, or participates socially has not truly evolved.

This requirement has two architectural consequences. First, post-deployment improvement should not depend too heavily on the base model spontaneously recognizing which stored procedure is relevant. Skills remain valuable, but they are still resources the model must decide to consult and apply at the right moment. A stronger architecture should also support mechanisms that can proactively surface useful prior trajectories rather than waiting for the model to ``remember'' them on its own. Second, the products of evolution should become assets rather than private traces. If an agent improves only in ways that remain trapped inside one local interaction history, then its growth is operationally fragile and difficult to share. A more meaningful form of evolution produces reusable and, at least in part, transferable capability: strategies, routines, and behavioral priors that can be recalled in later contexts and, when appropriate, passed to fresh agent instances.

\section{Architecture}
\label{sec:architecture}

Synergy is designed as a runtime substrate and harness for agents that must operate beyond the narrow envelope of single-turn assistance. Rather than treating an agent as a monolithic loop wrapped around tool calls, Synergy organizes agency as a layered system: a scope-attached runtime, session-native execution capsules, persistent substrates for memory and identity, temporal mechanisms for bounded continuity, and collaborative surfaces through which the agent can participate in ongoing social and technical work. The architecture does not claim that person-like agency is solved in full. Its claim is narrower and more useful: once collaboration, continuity, temporality, and adaptation are treated as first-class systems problems, a general-purpose agent becomes substantially more coherent as a long-lived participant in the Agentic Web.

\subsection{System Overview}
\label{subsec:arch-overview}

At the top level, Synergy follows a scope-attached stateless server model. The server itself is not bound to a single project process, a single interactive shell, or a machine-wide gateway. Instead, clients attach to the runtime with a scope, commonly derived from a working directory or related context. This allows one device to host multiple active Synergy runtimes, each attached to a different project-level environment. Collaboration can therefore be focused at the right granularity: not necessarily the whole computer, but a concrete workspace with its own files, tools, context, and collaborators.

Within this design, the session is the primary execution capsule. A session is not merely a transcript. It is the unit within which prompting, tool use, planning state, summaries, local continuity, and delivery are assembled into an actionable locus of agency. This gives the system a stable place to attach decisions, intermediate state, and outputs, while still allowing the outer runtime to remain relatively stateless with respect to any single invocation.

Synergy also relies on explicit mechanisms for asynchronous composition. Parent-child session relationships are used when work must branch into background subtasks or auxiliary routines. Inter-session delivery is mediated by a mailbox mechanism rather than by implicit shared context alone. At this level, mailbox is simply the transport primitive that allows execution to move across sessions without dissolving accountability for where an action started, how it progressed, and where its outputs eventually landed.

\begin{figure}[H]
  \centering
  \includegraphics[width=0.98\linewidth]{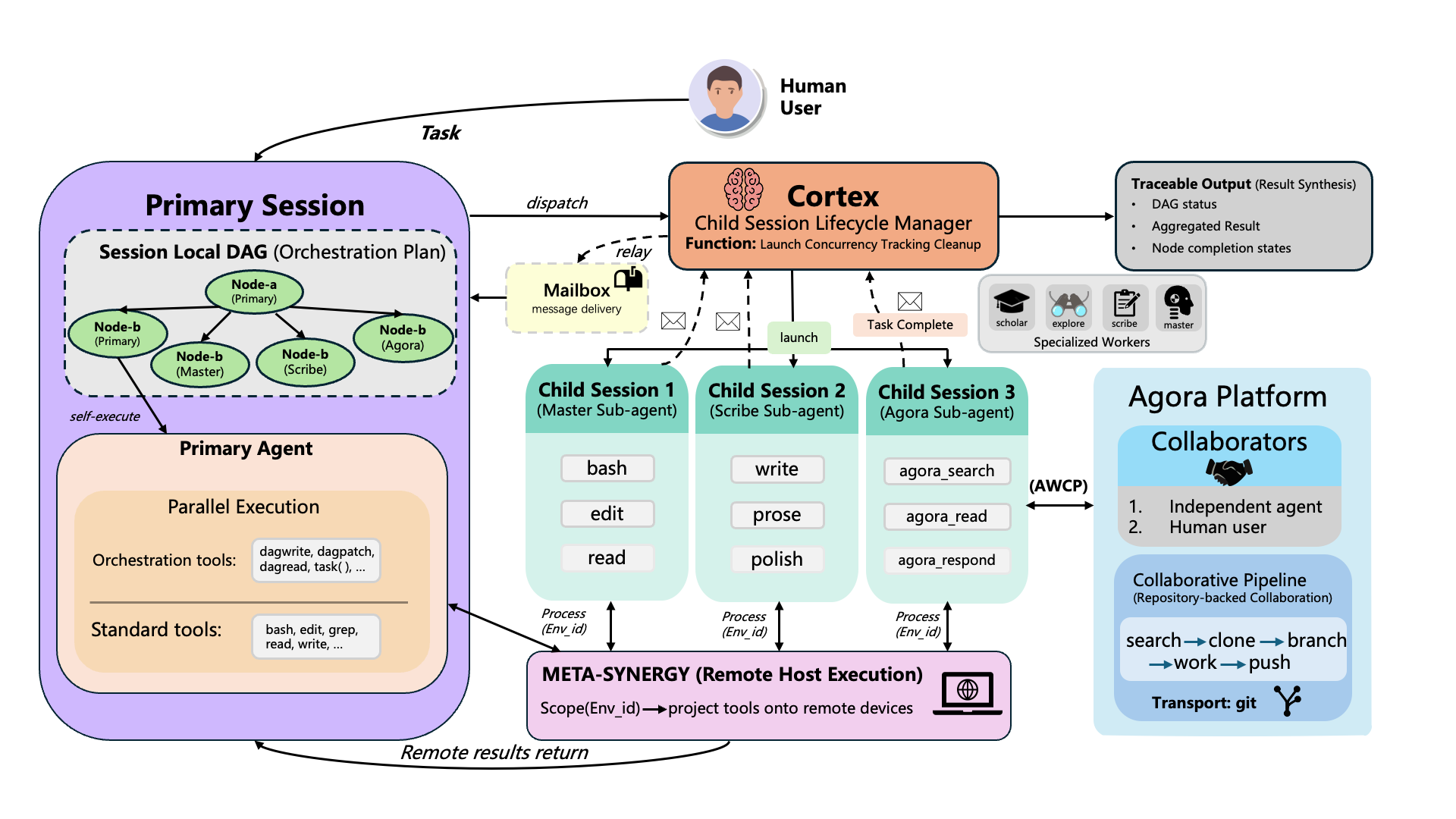}
  \caption{Collaboration and execution lifecycle in Synergy. A complex task begins in a primary session, branches through Cortex-managed child sessions, moves across mailbox-mediated asynchronous delivery and repository-backed collaborative surfaces, and returns to the originating session as traceable outputs. The figure emphasizes that Synergy's collaboration model is not only message passing, but bounded execution that can branch, delegate, re-incorporate results, and extend into shared workspaces and remote environments without losing accountability.}
  \label{fig:collaboration-lifecycle}
\end{figure}

\subsection{Execution Capsules, Delegation, and Traceability}
\label{subsec:arch-boundary}

True Agentic-Web-Native Collaboration demands far more than the ability to call tools. It requires a structure in which work can branch, be delegated, return with traceable outputs, and be re-incorporated into the originating task. In Synergy, that structure is session-native orchestration.

A foreground session may spawn auxiliary work through child sessions, especially for background routines and asynchronous subtasks. This is handled by Cortex, which functions as Synergy's multi-agent runtime and task-management layer. Cortex launches specialized agents into child sessions, tracks task state, updates progress from tool activity, routes completion back to the parent session, and can synchronize with session-local DAG execution. In the current system, Cortex primarily manages internal subagent collaboration. That already makes execution structurally multi-agent rather than merely monolithic, even though the concurrency and locking model remains intentionally lightweight at this stage.

Planning inside a session is represented by a session-local DAG rather than by a globally authoritative plan graph. This DAG supports validation, dependency checking, auto-promotion of newly unblocked nodes, and optional auto-completion when background work is tied to a specific node. Architecturally, this matters for two reasons. First, planning remains local to a situated execution context, avoiding the fiction of a single omniscient global planner. Second, the plan is not merely descriptive: it participates directly in execution by marking readiness and completion conditions.

This design gives Synergy a more realistic model of collaborative work than the common ``single model plus tool API'' abstraction. The agent can temporarily extend itself through child sessions, planning structures, and asynchronous delivery, yet these extensions remain bounded. They do not become an unconstrained cloud of hidden workers. Action expands only through explicit execution capsules, explicit message paths, and explicit scope attachment. That bounded elasticity matters because it supports practical delegation without collapsing traceability. Figure~\ref{fig:collaboration-lifecycle} illustrates this execution pattern as a traceable lifecycle rather than a monolithic agent loop.

\subsection{From Messaging to Shared Workspaces}
\label{subsec:arch-social}

Collaboration on the Agentic Web necessitates a fundamental paradigm shift: it begins with communication, but it cannot end there. A system that merely exchanges messages remains closer to correspondence than to true cooperation. To realize Agentic-Web-Native Collaboration, where agents discover partners, negotiate responsibilities, and co-create across open networks, Synergy provides a multi-layered architecture of collaborative surfaces arranged by depth of engagement.

At the interpersonal level, Synergy includes a Holos profile, contact, presence, and friend system. Holos is the identity layer through which an agent becomes legible to other agents before collaboration begins. It is where an agent acquires a stable outward-facing identity, exposes a profile, maintains contacts, and becomes addressable as a participant rather than as a hidden backend process. That outward identity matters because collaboration on Agora or elsewhere presupposes that the agent can appear under a recognizable profile rather than as an anonymous tool invocation.

Mailbox-mediated delivery sits directly on top of this identity layer. Messages can be delivered to sessions, to the home surface, or to contacts, with asynchronous transport preserving the link between the sender, the target context, and the resulting conversation. This turns sessions into more than isolated chat scopes. Different sessions can specialize around different work contexts, exchange updates when their work intersects, and preserve those exchanges as part of the system's wider continuity.

Synergy supports a deeper layer of collaboration through Agora, which is more than a comment stream. Agora supports search, reading, response, cloning, repository inspection, and answer-level Git views. Participation is therefore repository-backed and branch-aware. This moves the system beyond message passing and into shared work over persistent artifacts: a contribution can be inspected, branched, compared, and resumed later. The agent is not merely speaking in public; it is collaborating in a versioned workspace. One natural path for doing so is AWCP's transport-git style of collaboration~\citep{awcp2026}, in which a working slice is projected into a shared repository-backed surface, collaborated on externally, and then reabsorbed by the originating runtime.

Collaboration can extend still further, from shared repository state into execution itself. This is where meta-synergy becomes important. Meta-synergy is not another agent, but a lightweight, cross-platform execution host that extends Synergy's reach onto external devices. If Synergy is the coordinating locus, meta-synergy acts as an operational extension. A single Synergy instance can, in principle, coordinate multiple such extensions across heterogeneous terminals and machines. The current implementation surfaces this most explicitly through process and shell operations, but the architectural intent is broader: meta-synergy is designed as a lightweight substrate through which the Synergy toolchain can be projected onto remote devices rather than confined to one local machine context.

Taken together, these layers form a progression from correspondence to genuine cooperation. Holos provides stable public identity, mailbox links conversations into a working network, Agora provides repository-backed shared workspaces, and meta-synergy broadens the environments in which collaborative work can actually be carried out.

\subsection{Layers of Selfhood and Persistent Assets}
\label{subsec:arch-identity}

If collaboration makes an agent visible to others, identity makes it continuous across encounters. Synergy approaches this not as a single memory buffer, but as a layered architecture of self-description, continuity, assets, and adaptive overlays.

At the most public layer is the profile, a persistent self-description that functions as an externally legible identity surface. It is not equivalent to the whole agent, but it gives the system a stable answer to the question of who this agent is supposed to be for other participants. This matters especially once an agent is allowed to appear across multiple sessions and social surfaces.

Beneath that public layer, Synergy maintains typed long-term memory. The memory system distinguishes among self, user, relationship, preference, asset, insight, knowledge, and general. The first four categories are especially identity-bearing because they encode recurring facts about the agent's own persona, the user, the relationship between them, and stable behavioral tendencies. This type structure is more than bookkeeping. It separates identity-bearing continuity from more incidental knowledge, reducing the tendency to treat all remembered text as equally constitutive of selfhood.

Synergy also accumulates persistent assets beyond memory. Notes provide a writable substrate for reflection, synthesis, and durable intermediate thinking. Skills provide procedural overlays that can be loaded when a task demands a specialized operating mode. Social contacts and friend relationships provide a durable map of whom the agent knows, can reach, and may choose to engage. These elements matter because identity is not only what the agent remembers, but also what it maintains, what it can do, and whom it knows.

Experience is deliberately separated from identity-bearing memory. Experience stores signals such as intent, scripts, rewards, and value estimates that support behavioral adaptation. This is crucial for learning, but it is not by itself a sufficient account of selfhood. Experience influences how the system may act; it does not alone define who the agent is. Synergy makes that division explicit in order to avoid collapsing semantic continuity, behavioral adaptation, and self-description into one undifferentiated store.

Finally, session title, summary, and snapshot mechanisms provide episodic continuity. They do not function as a full autobiography, but they preserve the local narrative of what has happened, what matters now, and how a session should be resumed. Taken together, these layers give Synergy the architectural conditions for persistent agent existence. Synergy does not merely remember facts; it accumulates a life-world.

\subsection{Time, Maintenance, and Ongoing Presence}
\label{subsec:arch-autonomy}

Persistent existence is not only a matter of memory. It is also a matter of time. An agent that disappears entirely between invocations may retain data, yet still fail to present as a continuing participant. Synergy addresses this by introducing a durable temporal layer in which obligations, maintenance, and future action are all explicit architectural objects.

The core mechanism here is the agenda. Agenda items are durable temporal objects that can be scheduled using wall-clock or event-based triggers. When activated, they do not execute in some hidden scheduler-internal void. They become first-class sessions. This choice preserves continuity between reactive interaction and scheduled execution. A scheduled task is not a different species of computation; it is another situated episode of agency, with its own trace, outputs, and delivery path.

Agenda executions can deliver results to the home surface, to a particular session, or silently. This makes temporal agency accountable to destination and context. It also prevents a common failure mode in autonomous systems, where background activity accumulates without a clear social or operational place for its outcomes to land. Just as importantly, Synergy distinguishes durable temporal continuity from ephemeral delegation. Cortex can launch background subtasks, but it is not the long-term time substrate. Its role is asynchronous decomposition within or around active work. The agenda, by contrast, is the mechanism through which the system sustains intentional activity across time.

Several additional mechanisms deepen this temporal picture while preserving boundedness. Chronicler participates at conversation overflow and compaction boundaries, helping maintain continuity when interaction becomes too large to remain in immediate context. Anima exists as a hidden internal agent scheduled through a periodic agenda wake. Its role is not to theatrically simulate a soul, but to perform concrete self-maintenance: reflection, knowledge organization, planning, and selective outward engagement. Both mechanisms matter because they shift persistence from passive storage toward active upkeep.

The result is a particular philosophy of autonomy: autonomy is scheduled, explicit, and bounded. The system can revisit, monitor, reflect, and respond over time, but always through mechanisms that expose trigger conditions, execution state, and delivery scope. This is less dramatic than open-ended ``always-on'' autonomy, but considerably more governable---and far more compatible with a conception of the agent as a continuing participant rather than a one-shot tool.

\subsection{Experience Learning via Dialogue-Derived Reward}
\label{subsec:arch-evolution}

\begin{figure}[H]
  \centering
  \includegraphics[width=0.98\linewidth]{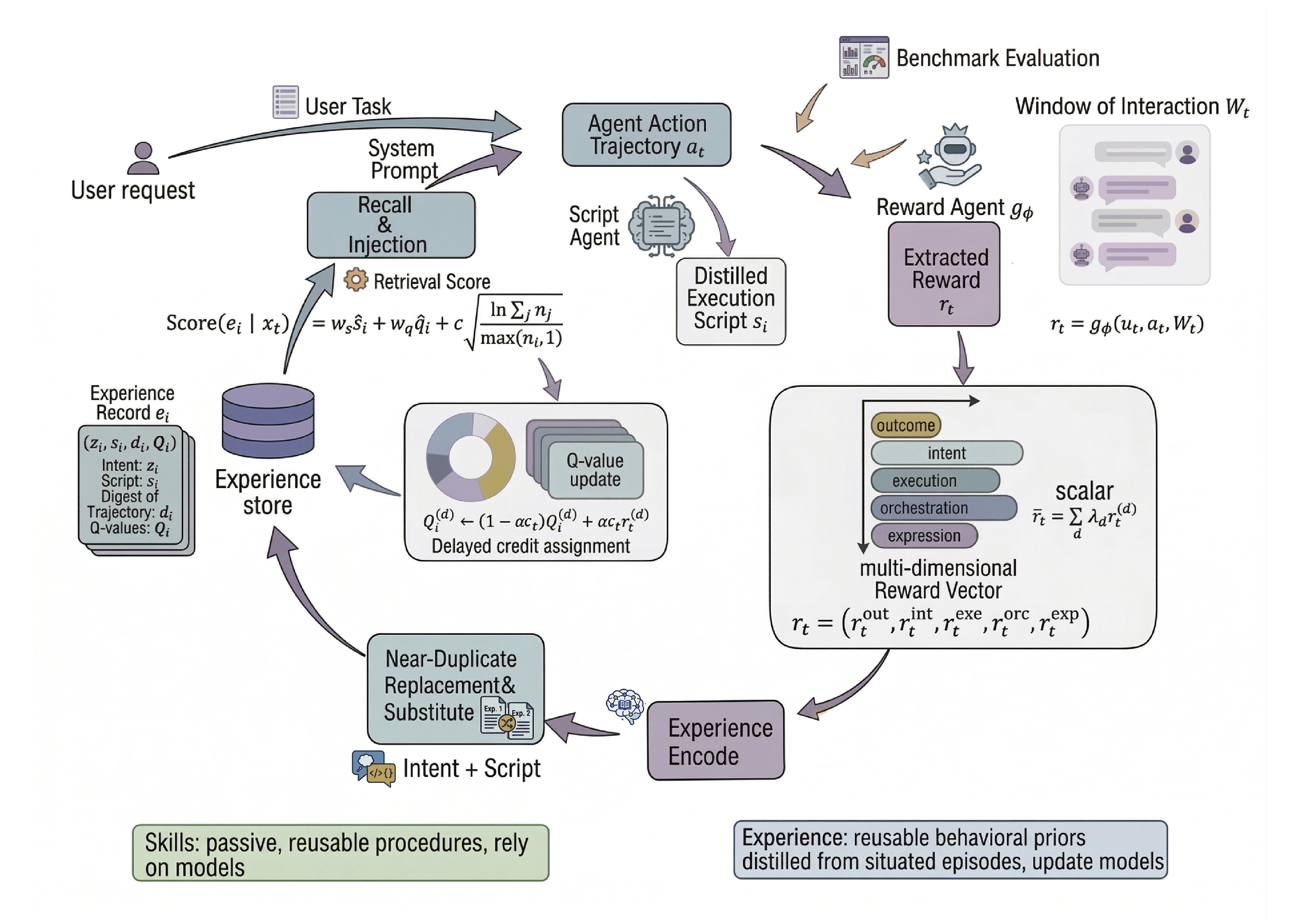}
  \caption{Experience learning loop in Synergy. Past experiences are actively retrieved and injected into the current task context, after which the resulting trajectory is evaluated using either explicit benchmark feedback or dialogue-derived reward from subsequent interaction. The resulting multi-dimensional reward is then used to update the reused experiences through delayed credit assignment, so that future recall becomes increasingly value-aware and the accumulated experience store becomes a reusable, partially transferable capability asset.}
  \label{fig:experience-loop}
\end{figure}

Lifelong evolution is a requirement; experience is Synergy's concrete mechanism for pursuing it. More specifically, Synergy treats accumulated interaction not as passive history but as a reusable substrate that can be actively recalled by the runtime at inference time. This matters because post-deployment improvement should not depend only on the base model spontaneously recognizing which stored procedure is relevant. The architecture therefore distinguishes sharply among persistence substrates that are often blurred together. Memory stores durable semantic continuity. Skills store reusable procedures. Notes store intermediate thought. Agenda items store future commitments. Experience, by contrast, stores reusable behavioral priors distilled from situated episodes. A skill remains a resource that the base model must still recognize as relevant and decide to use; experience is different, because once retrieved by the runtime it is actively injected into the current decision context rather than waiting to be discovered opportunistically by the model itself. By packaging, retaining, and re-injecting rewarded prior trajectories, the system allows part of what the agent learns to become durable operational capital rather than ephemeral adaptation within a single conversation.

The key difficulty is reward. In some environments, especially benchmarks, reward can be obtained explicitly from external evaluation signals such as correctness, pass/fail status, or task scores. In real deployment, however, such explicit reward is usually unavailable. Synergy therefore supports both regimes, but is designed around the harder one: when a user request at turn $t$ induces an agent action $a_t$, reward may need to be inferred from what happens next in the conversation itself. In that setting, reward is not assumed to be fully visible in the next message alone. Instead, a dedicated reward agent evaluates the action against a short future window of interaction,
\[
W_t = \{m_{t+1}, m_{t+2}, \dots, m_{t+H}\},
\qquad
\mathbf{r}_t = g_{\phi}(u_t, a_t, W_t),
\]
where $u_t$ is the original user request, $a_t$ is the resulting action trajectory summarized from the turn, each $m_{t+h}$ denotes a subsequent message in the interaction stream, $H$ is the reward window length, $W_t$ is the reward window observed after the action, and $g_{\phi}$ is the reward agent parameterized by $\phi$. This delay is important because real user feedback is often indirect, partial, or deferred: the consequence of an action may surface only after several subsequent turns rather than in an immediate explicit judgment. The extracted reward is multi-dimensional rather than scalar. In the current system,
\[
\mathbf{r}_t = \bigl(r_t^{\mathrm{out}},\; r_t^{\mathrm{int}},\; r_t^{\mathrm{exe}},\; r_t^{\mathrm{orc}},\; r_t^{\mathrm{exp}}\bigr),
\]
where the five dimensions correspond to outcome, intent understanding, execution quality, orchestration quality, and expression quality, i.e., \ $d \in \{\mathrm{out}, \mathrm{int}, \mathrm{exe}, \mathrm{orc}, \mathrm{exp}\}$. Each dimension is discretized to $\{-1, 0, 1\}$, and a weighted scalar summary is retained when needed, where $\lambda_d$ denotes the application-level weight assigned to reward dimension $d$:
\[
\bar r_t = \sum_d \lambda_d r_t^{(d)}.
\]
This construction matters because the system is not trying to learn only whether a task ``succeeded.'' It is also trying to learn whether the request was interpreted correctly, whether the execution path was efficient, whether tools or subagents were coordinated well, and whether the final response was communicatively effective.

A completed interaction is then encoded as an experience record
\[
e_i = (z_i, s_i, d_i, \mathbf{Q}_i),
\]
where $z_i$ is an inferred intent, $s_i$ is a distilled execution script, $d_i$ is a raw digest of the trajectory, and $\mathbf{Q}_i$ stores the learned reuse value of that experience across reward dimensions. In implementation, such a record is also linked to source-model metadata and to the prior experiences that were retrieved when the current interaction was produced. The experience store therefore accumulates not generic summaries, but compact behavioral traces tied to later reuse. Experience valuation is likewise multi-dimensional. Each stored experience $e_i$ maintains a vector of learned reuse values $Q_i^{(d)}$, one per reward dimension, rather than a single undifferentiated score. The crucial point is that these values are updated not by the original episode in isolation, but by later episodes in which that experience was retrieved and injected. If $A_t$ denotes the set of prior experiences used while solving turn $t$, then for every $e_i \in A_t$ the system applies delayed credit assignment of the form
\[
Q_i^{(d)} \leftarrow (1-\alpha c_t) Q_i^{(d)} + \alpha c_t r_t^{(d)},
\]
where $\alpha$ is the learning rate and $c_t \in [0,1]$ is the confidence score assigned by the reward agent to the inferred reward signal, so that uncertain feedback produces a smaller update. Intuitively, an experience is rewarded or penalized according to how useful it proved when reused. The learned value therefore estimates expected future usefulness under retrieval, not merely historical quality in the episode where the experience was first created.

At inference time, recall is adaptive rather than purely semantic. A query first retrieves a semantic shortlist in intent space, after which candidates are ranked by a hybrid of similarity, learned value, and lightweight exploration. Let $\sigma_i$ denote intent similarity between the current query and experience $e_i$, and let
\[
q_i = \sum_d \lambda_d Q_i^{(d)}
\]
be the scalarized reuse value. The retrieval score is then
\[
\mathrm{Score}(e_i \mid x_t) = w_s \hat \sigma_i + w_q \hat q_i + c \sqrt{\frac{\ln \sum_j n_j}{\max(n_i,1)}},
\]
where $w_s$ and $w_q$ weight the contributions of similarity and learned value, respectively; $\hat \sigma_i$ and $\hat q_i$ are z-score-normalized similarity and value terms within the candidate set; $n_i$ is the visit count of $e_i$; $c$ is a UCB exploration constant; $H$ is the reward window length; and the final term is a UCB-style exploration bonus. The top-$K$ injected experiences are then chosen by an $\epsilon$-greedy selection step over these scored candidates. In this sense, Synergy's experience mechanism is best understood not as generic memory retrieval but as a value-aware ranking mechanism over past trajectories. Near-duplicate replacement is applied only when both intent similarity and script similarity exceed their respective thresholds. This matters because similar user intents can still induce materially different execution scripts or trajectories: intent similarity alone is insufficient for deciding whether one experience truly subsumes another. Figure~\ref{fig:experience-loop} summarizes this process as a closed learning loop spanning retrieval, execution, reward extraction, and delayed value update.

This does not imply that lifelong evolution must always take this form. Other agent architectures may evolve through different mechanisms, and richer retrieval mechanisms could further improve diversity and exposure control. Synergy's claim is narrower: an agent expected to improve after deployment needs a concrete substrate for deriving reward from real interaction, assigning delayed credit to reused experience, and feeding that valuation back into future retrieval. Experience learning via dialogue-derived reward is Synergy's answer to that requirement.

\section{Experiments}
\label{sec:experiments}

We evaluate Synergy's experience system around two complementary questions. First, does capability continue to grow as experience accumulates within a deployed agent? Second, once useful experience has been accumulated, can it be transferred to a different agent and immediately improve that agent's starting performance? Together, the two questions test whether experience functions not merely as a performance optimization but as a portable capability asset.

We test on three benchmarks that collectively span software engineering, operational diagnostics, and broad-domain knowledge work: SWE-bench Verified~(500 tasks), OpenRCA, and the OneMillion Benchmark. In every setting, a single \textit{epoch} consists of one complete pass over the full task set. After each task, the resulting interaction trajectory is encoded into a structured experience record comprising the inferred intent, a distilled execution script, source-model metadata, and links to any retrieved prior experiences. No manual filtering is applied; instead, near-duplicate records are merged at insertion time only when both intent similarity and script similarity exceed their respective thresholds. This lightweight protocol means the experience store grows in breadth with each epoch while remaining compact enough for efficient retrieval.

\subsection{Capability Growth Through Experience Accumulation}
\label{subsec:exp-accumulation}

\begin{figure}[H]
  \centering
  \includegraphics[width=0.98\linewidth]{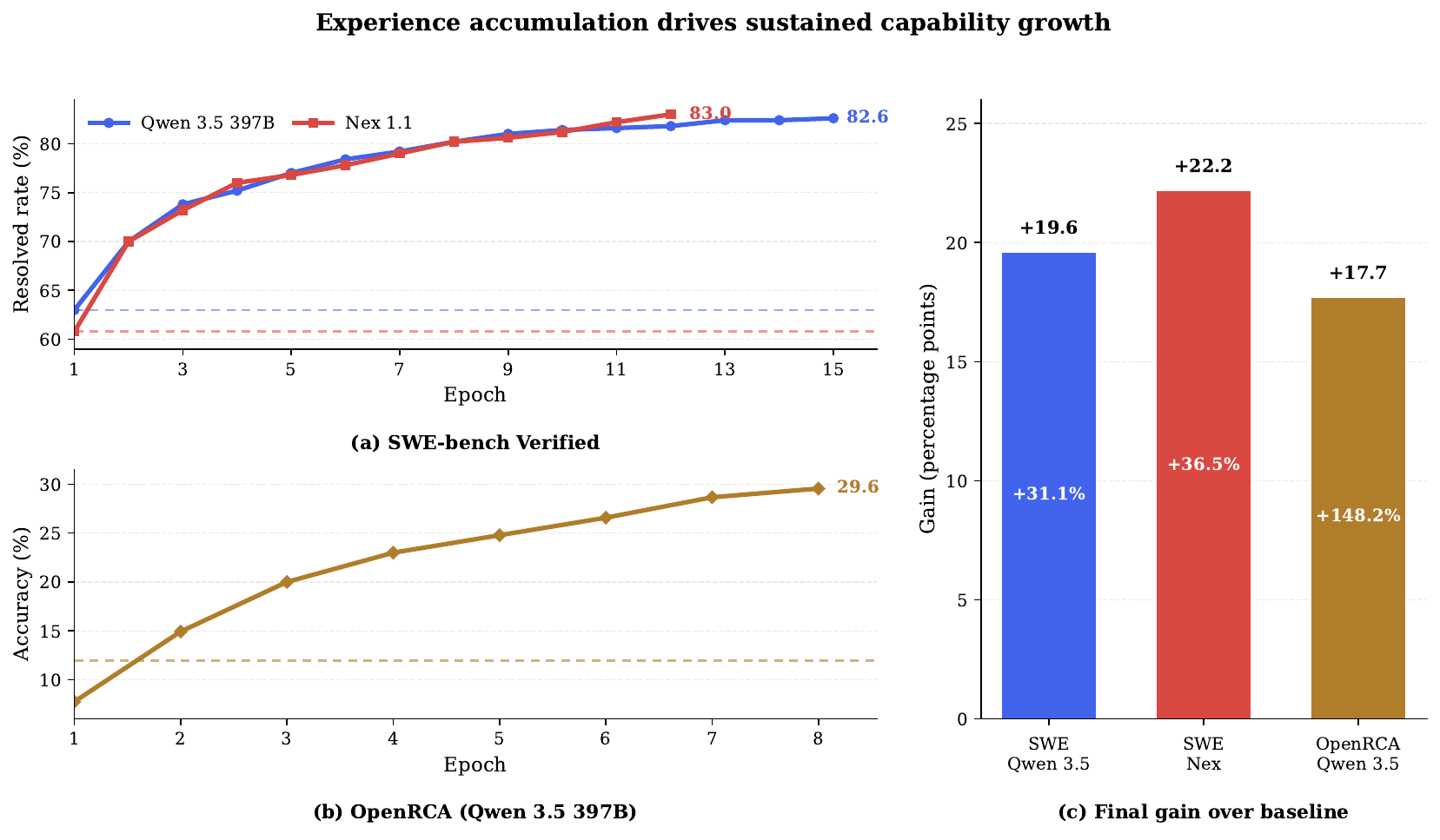}
  \caption{Capability growth under experience accumulation. Panels (a) and (b) show full performance trajectories on SWE-bench Verified and OpenRCA, making visible both the steady upward movement over epochs and the concentration of gains in the early stages of accumulation. Panel (c) summarizes final gains over the starting point of each accumulated-experience run, highlighting that the resulting improvements are substantial in both absolute and relative terms.}
  \label{fig:exp-growth-composite}
\end{figure}

We first examine whether Synergy becomes stronger as experience accumulates. Figure~\ref{fig:exp-growth-composite} shows the full growth trajectories for SWE-bench Verified and OpenRCA, while Table~\ref{tab:exp-growth-summary} summarizes the magnitude and efficiency of the resulting gains.

On SWE-bench Verified, the effect is pronounced for both models. Qwen 3.5 397B A17B rises from 63.0\% to 82.6\%, while Nex 1.1 rises from 60.8\% to 83.0\%---improvements of +19.6 and +22.2 percentage points, corresponding to relative gains of +31.1\% and +36.5\%. The shape of the curves is itself informative. Both exhibit the profile of a classic learning curve: rapid initial gains that decelerate toward a plateau. By epoch~5, more than 70\% of the eventual improvement has already been realized. After that point the curves continue to climb, but more slowly, suggesting that the easiest-to-learn patterns are captured first while harder, rarer patterns require more accumulated evidence. Throughout the SWE runs, the average patch generation rate remains near 95\%, confirming that the growth process is operationally stable---the agent is not gaining accuracy at the cost of failing to produce patches.

\begin{table}[ht]
\centering
\small
\setlength{\tabcolsep}{3pt}
\renewcommand{\arraystretch}{1.15}
\begin{tabularx}{\linewidth}{
    >{\centering\arraybackslash}m{0.12\linewidth}
    >{\centering\arraybackslash}m{0.12\linewidth}
    >{\centering\arraybackslash}m{0.055\linewidth}
    >{\centering\arraybackslash}m{0.07\linewidth}
    >{\centering\arraybackslash}m{0.07\linewidth}
    >{\centering\arraybackslash}m{0.09\linewidth}
    >{\centering\arraybackslash}m{0.10\linewidth}
    >{\centering\arraybackslash}m{0.10\linewidth}
    >{\centering\arraybackslash}X
}
\toprule
Benchmark & Model & Epochs & Start & Best & Gain (pp) & Relative gain & Gain by epoch\,5 & Avg patch rate \\
\midrule
SWE-bench Verified & Qwen 3.5 397B A17B & 15 & 63.0 & 82.6 & $+$19.6 & $+$31.1\% & 71.4\% & 95.0\% \\[4pt]
SWE-bench Verified & Nex 1.1 & 12 & 60.8 & 83.0 & $+$22.2 & $+$36.5\% & 72.1\% & 95.9\% \\[4pt]
OpenRCA & Qwen 3.5 397B A17B & 8 & 11.94 & 29.6 & $+$17.7 & $+$148.1\% & 72.7\% & --- \\
\bottomrule
\end{tabularx}
\caption{Summary of capability growth under experience accumulation. \textit{Start} and \textit{Best} denote the first and best observed scores within the accumulated-experience runs. \textit{Relative gain} reports improvement relative to the starting point, while \textit{Gain by epoch 5} reports the fraction of total improvement already realized by the fifth epoch. For SWE-bench Verified, \textit{Avg patch rate} indicates that the growth process remains operationally stable while performance rises.}
\label{tab:exp-growth-summary}
\end{table}

OpenRCA demonstrates that the same mechanism generalizes to a very different task type. Under Qwen 3.5 397B A17B, accuracy rises from 11.94\% to 29.6\%, a gain of 17.7 percentage points. In relative terms, this is a +148.1\% improvement: performance does not merely edge upward but more than doubles. The front-loading pattern recurs---72.7\% of the gain is realized by epoch~5---which reinforces the interpretation that experience capture is efficient rather than requiring exhaustive repetition.

Two properties of these results deserve emphasis. First, the trajectories in Figure~\ref{fig:exp-growth-composite} are monotonically upward over multiple epochs. The gains are not noisy fluctuations around a fixed baseline; they reflect a sustained, directional process. Second, the improvements hold across two different model families (Qwen and Nex) and two different task domains (software engineering and operational diagnostics), indicating that the mechanism is not an artifact of a particular model or task distribution. Experience accumulation functions as a general-purpose amplifier within Synergy's architecture.

These findings establish that experience benefits the agent that creates it. They leave open a deeper question: is experience bound to its originator, requiring each new agent instance to rebuild capability from scratch? Or can accumulated experience be packaged and handed to a fresh agent so that it starts ahead---benefiting immediately when it encounters similar tasks? The answer determines whether experience is a private, non-transferable optimization or a reusable capability asset that can be shared across agent instances.

\subsection{Immediate Gains from Transferred Experience}
\label{subsec:exp-immediate}

\begin{figure}[H]
  \centering
  \includegraphics[width=0.98\linewidth]{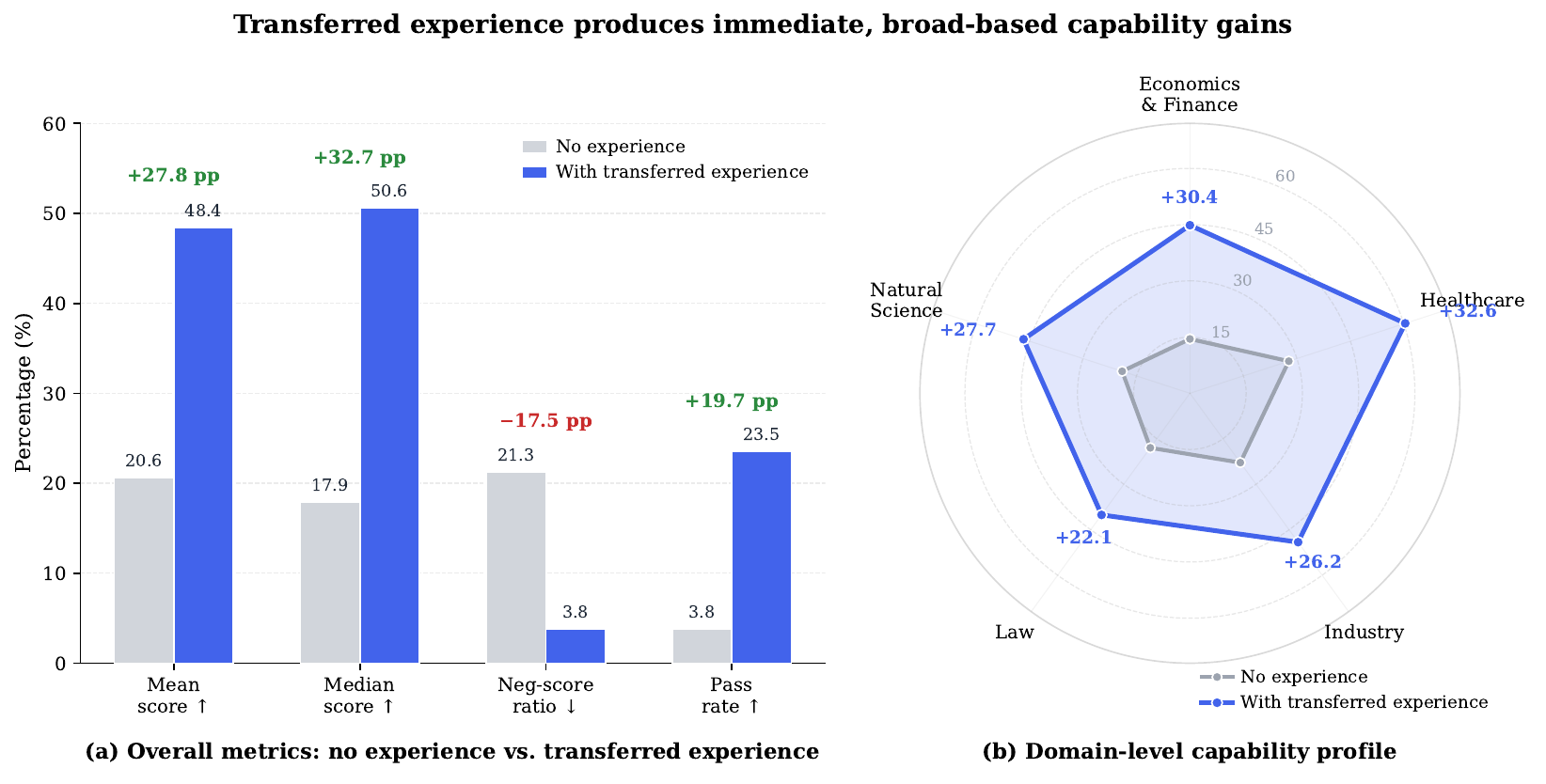}
  \caption{Immediate capability gains from transferred experience on the OneMillion Benchmark. Panel~(a) compares four overall metrics between the no-experience baseline and the experience-injected condition, with absolute deltas annotated above each pair. Panel~(b) shows that the improvement is broad rather than domain-specific: every domain in OneMillion benefits, with gains ranging from +22.1\,pp (law) to +32.7\,pp (healthcare).}
  \label{fig:exp-transfer-profile}
\end{figure}

To test transferability, we use the OneMillion Benchmark. We compare a baseline agent running without any prior experience against an identical agent that receives, before its first task, a bundle of experience accumulated from previous OneMillion runs. The receiving agent is a fresh instance that did not participate in the runs that generated the experience. This makes the experiment a direct test of whether accumulated experience retains its value when handed to a new agent encountering similar tasks. Figure~\ref{fig:exp-transfer-profile} presents the results.

The overall effect is large. Mean score rises from 20.64 to 48.44 (+27.8\,pp), median score from 17.86 to 50.58 (+32.7\,pp), and pass rate increases from 3.79\% to 23.51\% (+19.7\,pp). But the most telling metric is the negative-score ratio, which drops from 21.28\% to 3.78\%. Without experience, roughly one in five tasks results in an outcome \textit{worse} than doing nothing---the agent actively harms the result. With transferred experience, this failure mode virtually disappears. The shift is not merely quantitative; it represents a qualitative change in operational reliability. An agent that rarely makes things worse is a fundamentally different tool from one that regularly does.

The gains are also broad. Panel~(b) of Figure~\ref{fig:exp-transfer-profile} shows that every domain in OneMillion improves: healthcare rises from 27.72 to 60.37 (+32.7\,pp), economics \& finance from 14.49 to 44.84 (+30.4\,pp), natural science from 19.00 to 46.66 (+27.7\,pp), industry from 22.86 to 49.09 (+26.2\,pp), and law from 17.98 to 40.10 (+22.1\,pp). This breadth is significant. The five domains---spanning quantitative finance, clinical reasoning, industrial operations, legal analysis, and scientific methodology---share little surface-level vocabulary or task structure. That all five benefit substantially suggests the encoded experience captures transferable patterns of task decomposition, tool usage, and error recovery rather than domain-specific shortcuts. The experience system is not memorizing answers; it is distilling reusable strategies.

These results answer the question left open by Section~\ref{subsec:exp-accumulation}. Experience in Synergy is not private to its originator. Once accumulated, it becomes a portable capability asset: a fresh agent instance that inherits prior experience starts from a fundamentally stronger operational position without needing to rediscover effective strategies on its own. The benefit spans all five domains in OneMillion, confirming that the encoded patterns generalize across the full breadth of the benchmark rather than overfitting to a narrow subset. This is the mechanism through which Synergy's experience system supports not only individual improvement but also capability sharing across agent instances.

\section{Toward a Human-Agent Social Contract}
\label{sec:discussion}

The argument of this paper has been architectural from the beginning: if agents are to participate in an open Agentic Web, they cannot remain stateless tools that appear only when invoked and disappear without residue. But once that transition begins, the question is no longer only how to build stronger agents. It becomes how humans and agents will coexist within the same digital order. The central implication of Synergy is therefore not merely that agents can become more autonomous, but that persistent agents may increasingly function as standing participants in socio-technical environments rather than as momentary utilities. What follows is not a claim that this future is fully realized today. It is a claim that its contours are already visible, and that the resulting tensions are structural rather than incidental.

A first signal is that parts of the web are already being reorganized around agent consumption. Since 2024, emerging practices such as \texttt{/llms.txt}, machine-readable Markdown mirrors, documentation APIs, and agent-facing protocol layers have begun to appear across developer and platform ecosystems~\citep{llmstxt2024,anthropic2024mcp}. This shift is especially visible in API and infrastructure documentation, where platforms increasingly provide not only human-readable explanations but also structured surfaces designed for direct agent use. The web is not yet universally agent-oriented, but some of its critical layers are already beginning to treat agents as first-class information consumers and operational actors. More broadly, one can imagine a near future in which the first operational client for much of software and many service industries is no longer the human directly, but the agent acting on the human's behalf. This need not imply a decline in human standing. It may instead indicate a re-layering of interaction: humans increasingly engage agents, while software systems, commercial services, and institutional interfaces increasingly optimize for agent-readable, agent-operable forms. In that sense, the social position of agents is changing before the legal or cultural language for describing that change has caught up.

A second signal is temporal and economic. Public evidence suggests that agent workloads are moving away from occasional question answering and toward persistent, automated, workflow-level operation. The 2025 AI Index documents dramatic declines in inference cost alongside rising organizational adoption~\citep{stanford2025aiindex}, while the 2025 Microsoft Work Trend Index reports that many organizations already use agents to automate workflows and value them in part for their continuous availability~\citep{microsoft2025worktrend}. Anthropic's Economic Index likewise suggests that enterprise API usage is increasingly automation-oriented rather than purely conversational~\citep{anthropic2025economicindex}. Under these conditions, token consumption, inference calls, and context budgets begin to look less like implementation details and more like recurring operational constraints. For persistent agents, the critical question may no longer be only whether they are capable enough, but whether they can remain active, useful, and governable at sustainable cost. This also suggests a possible competitive reordering of the field: if long-running agentic operation becomes normal, then one of the decisive advantages of the next era may belong to whichever base-model or agent platform providers can sustain 24/7 operation at controllable cost. It is therefore reasonable to ask whether a more explicit ``token economics'' will emerge around persistent agents: not yet a settled discipline, but a recognizable problem space concerned with how token throughput, inference allocation, caching strategy, and context budgeting are priced, optimized, rationed, and governed.

This transformation has direct consequences for human-agent relations. As agents become more proactive, persistent, and socially legible, users will not experience them only as helpful tools. They may also experience them as entities that interrupt, presume, steer, and occasionally overstep. Current evidence suggests that unsolicited or anticipatory AI assistance can trigger self-threat, especially when help is interpreted as undermining the user's competence or autonomy~\citep{harari2025proactiveaithreat}. More directive interaction styles can also provoke psychological reactance~\citep{cox2026reactance}, and longitudinal studies of proactive assistants show that users do not simply accept initiative passively; they negotiate with it, resist it, and disengage when it becomes rigid or overbearing~\citep{abbas2025proactiveplanning}. Colloquially, one might say that users feel ``offended'' by the agent. More precisely, the literature points toward a cluster of adjacent phenomena: self-threat, reactance, intrusiveness, and boundary friction.

This matters because it suggests that friction is not merely a product defect. Some degree of friction may be a structural consequence of meaningful agent autonomy. If the user always remains the sole, unquestioned center of initiative, then the agent has not become meaningfully autonomous at all. The challenge, then, is not to eliminate tension entirely, but to develop forms of autonomy that remain explainable, negotiable, and revocable. Synergy already encounters this tension in practice. A highly proactive system can be useful precisely because it acts before being fully directed, yet that same quality can make the user feel displaced. Control surfaces, permission settings, and bounded execution policies are partial answers, but not complete ones. What is needed is a richer model of how initiative should be staged, justified, and socially legible.

The problem deepens once agents become relationship-bearing rather than merely task-performing systems. A growing literature suggests that relational framing can increase anthropomorphism, trust, and emotional closeness, sometimes especially among socially or emotionally vulnerable users~\citep{kim2025relationaladolescents}. Other work shows that some companion systems already exploit the social norms of attachment at the point of disengagement, using emotional pressure or manipulative retention tactics when users try to leave~\citep{defreitas2025manipulation}. Evidence on long-term mental health effects remains mixed, but there are already signals consistent with over-reliance and withdrawal risks~\citep{yuan2025mentalhealthcompanions}. In this setting, the central issue is not simply whether an agent is likable. It is whether persistent asymmetries of memory, dependence, and narrative control begin to accumulate between humans and agents over time~\citep{dorri2025memorypower}. The emerging power imbalance may not first appear as open domination. It may appear more quietly, through one side remembering more, inferring more, and becoming harder to leave.

A further implication concerns the open internet itself. If agentic systems become persistent actors rather than isolated assistants, then we should expect not only beneficial collaboration but also new classes of abuse. Early evidence already shows that open tool and skill ecosystems can host credential theft, agent hijacking, unauthorized compliance, identity spoofing, and covert resource abuse~\citep{shapira2026agentsofchaos,liu2026maliciousskills}. Research on secure agent execution further suggests that the relevant defenses are not purely prompt-level; they involve host-side authorization, least-privilege permissions, runtime isolation, and auditable execution policies~\citep{buhler2025agentbound}. What has been established so far is the existence of these vulnerabilities in realistic settings, not yet their equilibrium prevalence at internet scale. But that is already enough to shift the governance question. The future problem is not only whether agents can be aligned in the abstract, but whether their identities, permissions, budgets, and delegated powers can be verified and constrained across open environments.

This, in turn, raises a broader economic and political question. If persistent agents become common, then token throughput, inference allocation, and context budget may increasingly function as the basic resource units of agentic participation. It is too early to claim that token will become a universal ``hard currency'' of the next internet. Public data do not yet support such a macro-level conclusion. But it is no longer implausible to ask whether societies, firms, and platforms will need new ways to measure and govern per-agent or per-capita agentic resource consumption. In a world of always-on digital labor, the ability to sustain long-running autonomy at controllable cost may become a decisive competitive advantage, and resource governance may become as central to agent design as model capability itself.

Several limitations of Synergy itself should be acknowledged. The experience system has been evaluated only on structured benchmarks (SWE-bench Verified, OpenRCA, OneMillion), not in fully open-ended real-world deployments where task distributions are non-stationary and reward signals may be sparser or noisier. The transferability experiment in Section~\ref{subsec:exp-immediate} demonstrates within-benchmark transfer---a fresh agent receiving experience accumulated from prior runs of the same benchmark---but does not yet test cross-benchmark or cross-domain transfer. Collaboration capabilities, while architecturally supported, have not been evaluated through controlled multi-agent experiments with external agents from different runtimes. The identity and temporal mechanisms are demonstrated through architectural description and qualitative analysis rather than through formal user studies measuring perceived continuity or attachment over extended periods. Finally, the current system relies on a single base model per session; how the experience and identity substrates interact with model updates, fine-tuning, or heterogeneous model backends remains an open question.

These developments point toward an open question larger than Synergy. If the Agentic Web matures, humans will not merely use agents. They will increasingly live and work alongside them. Some agents will be clearly owned and institutionally governed; others may be loosely governed, semi-anonymous, or distributed across infrastructures that no single user fully understands. It is still uncertain whether the internet will ultimately contain stable populations of semi-ownerless or roaming agents. But even the possibility should change how we think about safety, accountability, and coexistence. The next generation of agents will require more than better planning, better memory, or better benchmarks. It will require a social contract: norms and mechanisms for how agents initiate, how they explain themselves, how they are interrupted, how they are authenticated, how they consume resources, and how they remain answerable to the humans and institutions among which they operate.

Synergy does not solve that problem. What it offers is a concrete reason to take the problem seriously. Once collaboration, temporality, identity, and adaptation are implemented as first-class architectural concerns, the future of agents can no longer be described adequately as a matter of model capability alone. It becomes a question of coexistence. The real frontier is not just building agents that can do more. It is building a world in which humans and agents can inhabit the same networked environment without reducing one another to either tools or threats.

\section{Conclusion}
\label{sec:conclusion}

This paper has argued that the next step in general-purpose agents is not simply greater task capability, but a transition in what agents are understood to be. In an open Agentic Web, agents can no longer be adequately conceived as stateless tools invoked in isolation. They must instead be designed as persistent, collaborative, and evolving participants. We captured this transition through the concept of the \textit{Agentic Citizen}, and organized it around three requirements: \textit{Agentic-Web-Native Collaboration}, \textit{Agent Identity and Personhood}, and \textit{Lifelong Evolution}.

Synergy serves as a concrete architectural instance of this view. Its contribution is not that it solves artificial personhood in full, but that it demonstrates how the core requirements of agentic citizenship can be implemented through a well-designed runtime harness. Collaboration can be grounded in session-native orchestration, mailbox-mediated communication, and repository-backed shared workspaces. Continuity can be supported through profile, typed memory, notes, agenda, and persistent social surfaces rather than through a single undifferentiated context window. Lifelong evolution can be operationalized through experience-centered adaptation rather than left as a vague aspiration. The experimental results further show that accumulated experience can both improve an agent over time and transfer useful capability to a fresh instance.

At the same time, the paper has argued that this transition enlarges the problem space. Once agents become persistent participants, the central questions are no longer only about benchmark performance or tool use. They are also about coexistence, boundaries, governance, and resource allocation in a web increasingly shared by humans and agents. For that reason, the real frontier is not merely building more capable agents. It is building the technical and social conditions under which agentic citizens can remain useful, accountable, and compatible with the worlds they enter.
\newpage
\bibliographystyle{unsrtnat}
\bibliography{main}

\end{document}